\documentclass[
letterpaper,
amsmath,
aps,
prl, twocolumn,reprint,showpacs,
]{revtex4-1}

\usepackage{hyperref}
\usepackage[english]{babel}
\usepackage[utf8]{inputenc}
\usepackage{graphicx}
\usepackage{epstopdf}
\usepackage[colorinlistoftodos]{todonotes}
\newcommand{\br}{{\bf r}}
\newcommand{\bn}{{\bf n}}

\newcommand{\bu}{{\bf u}}

\newcommand{\bq}{{\bf q}}
\newcommand{\bN}{{\bf N}}
\newcommand{\bR}{{\bf R}}

\date{\today}

\begin{document}

\title{Transport barriers to self-propelled particles in fluid flows}

\date{\today}

\author{Simon A. Berman$^1$, John Buggeln$^2$, David A. Brantley$^{1,3}$, Kevin A. Mitchell$^1$, Thomas H. Solomon$^2$}
\affiliation{$^1$Department of Physics, University of California, Merced, CA 95344 USA}
\affiliation{$^2$Department of Physics and Astronomy, Bucknell University, Lewisburg, Pennsylvania 17837 USA}
\affiliation{$^3$Lawrence Livermore National Laboratory, Livermore, CA 94550, USA}


\begin{abstract}
  We present theory and experiments demonstrating the existence of
  invariant manifolds that impede the motion of 
  microswimmers in two-dimensional fluid flows.  One-way barriers are
  apparent in a hyperbolic fluid flow that block the swimming of both
  smooth-swimming and run-and-tumble \emph{Bacillus subtilis} bacteria.
  We identify key phase-space structures, called
  swimming invariant manifolds (SwIMs), that serve as separatrices
  between different regions of long-time swimmer behavior. When
  projected into $xy$-space, the edges of the SwIMs act as one-way
  barriers, consistent with the experiments.
\end{abstract}

\maketitle

Dynamically defined transport barriers \cite{Ottino1990,Aref2017} impede the motion of passive particles in a wide range of fluids, from microbiological and microfluidic flows to oceanic,  atmospheric, and stellar flows. 
For steady and time-periodic flows, transport barriers are identified
with invariant manifolds of fixed points and Kolmogorov-Arnold-Moser
surfaces \cite{Mackay1984,Rom-Kedar1990,Meiss2015}.
More recently, these ideas have been extended to
aperiodic and turbulent flows
\cite{Voth2002,Shadden2005,Coulliette2007,Mathur2007,Haller2015}.
However, in many systems of fundamental and practical importance, the
tracers are {\it active} rather than passive.  Examples include
propagating chemical reaction fronts \cite{Cencini2003,Saha2013},
aquatic vessels \cite{Rhoads2013a}, and artificial and biological
microswimmers \cite{Torney2007,Khurana2011}, including Janus particles \cite{Ebbens2010,Katuri2018} and flagellated bacteria
\cite{Wioland2013,Rusconi2014}.

Invariant manifold theory has previously been extended to incorporate
propagating reaction fronts in a flow \cite{Mahoney2012,Mitchell2012,Mahoney2013,Mahoney2015a,Locke2018}.
This theory identifies analogs of passive transport barriers, called
{\it burning invariant manifolds} (BIMs), which are one-way barriers
to front propagation.
Experiments on front propagation in driven fluid flows \cite{Bargteil2012,Megson2015,Mahoney2015,Doan2018} demonstrate the physical significance of these theories.
Despite this success with reaction fronts, a comparable understanding for more general active systems is lacking.

\begin{figure}
\includegraphics[width=0.45\textwidth]{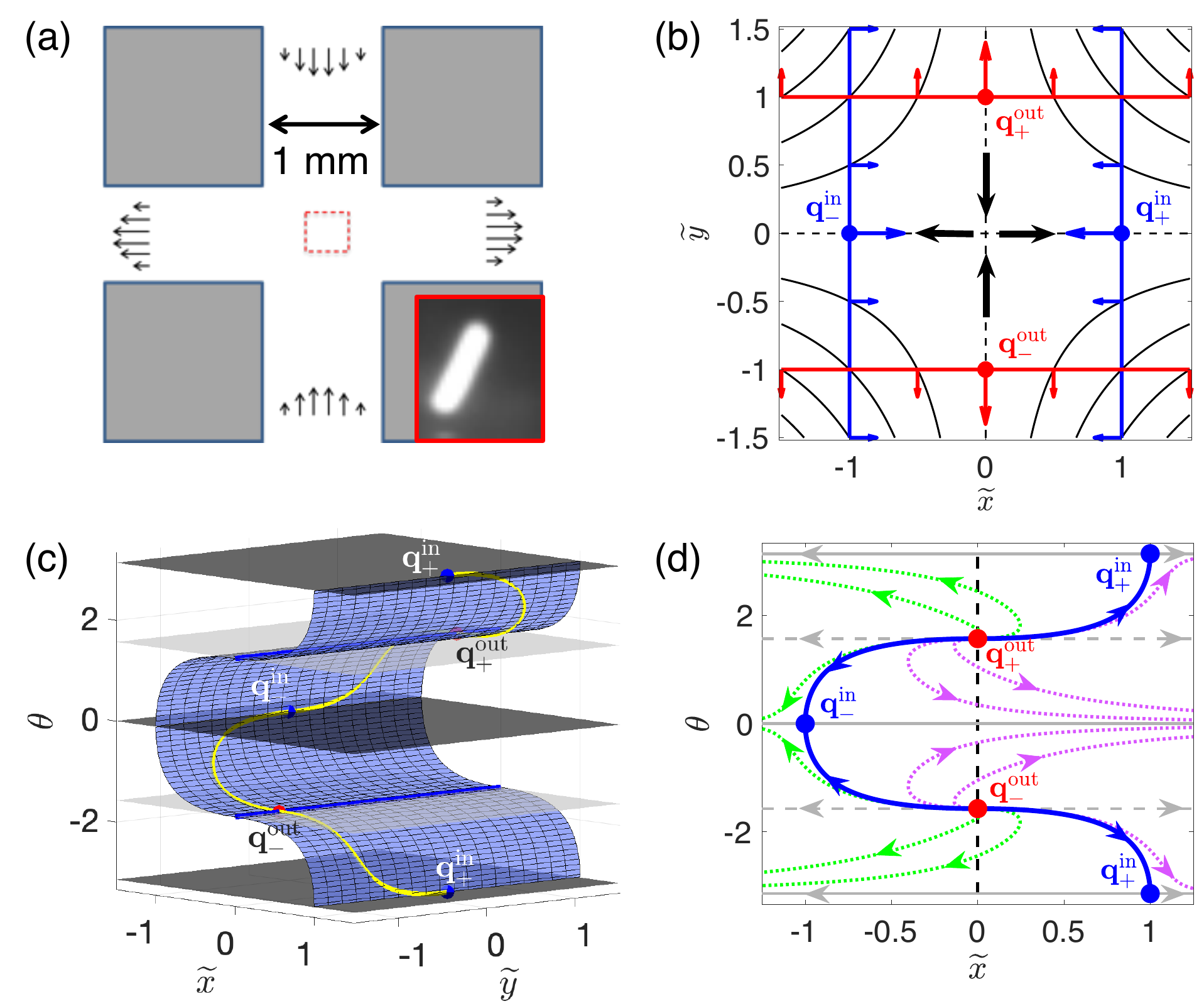}
\caption{(a) Cross-flow experiment; data is obtained in the red
square. Inset: 100X image of a fluorescent \emph{B.~subtilis}.
(b) SFPs and SwIM edges (in red/blue) of
the hyperbolic flow; $\alpha > 0$. Arrows indicate the direction of $\hat{\bn}$ (and the blocking direction) for the equilibria and the SwIM edges. Streamlines of the flow are plotted in black.
(c) Stable SwIMs (blue surfaces) of the $\bq^{\rm in}_{\pm}$ SFPs for $\alpha = 1$. The black (gray) planes are stable (unstable) invariant surfaces. The yellow curves are heteroclinic orbits connecting pairs of SFPs. (d) Constant $y$ cross-section of the swimmer phase space. The blue orbits are cross-sections of the stable SwIMs. 
}
\label{fig:hyperbolic_equilibria}
\end{figure}

This Letter presents theory and supporting experiments for a foundational and universal invariant manifold framework that describes barriers for active tracers in laminar fluid flows.
We focus on self-propelled particles, i.e.\ swimmers, and propose the existence of
\emph{swimming invariant manifolds} (SwIMs) that (i) act as absolute barriers
blocking the motion of smooth swimmers in position-orientation space;
(ii) project to one-way barriers in position space; and (iii) provide
insight into the motion of non-smooth (e.g. tumbling) swimmers.
We also find that (iv) one-way barriers exist even for tumbling swimmers, and these barriers turn out to be the BIMs that were previously shown to be barriers for reaction fronts \cite{Mahoney2012}.
Our experiments use smooth-swimming and
run-and-tumble strains of \emph{Bacillus subtilis} bacteria
(Fig.~\ref{fig:hyperbolic_equilibria}a inset) as active tracers in a
laminar, hyperbolic flow in a microfluidic cross-channel
(Fig.~\ref{fig:hyperbolic_equilibria}a).  Absent Brownian motion,
passive tracers in a linear hyperbolic flow cannot traverse the
passive invariant manifolds (separatrices) forming a cross along the
channel centerlines (dashed lines in
Fig.~\ref{fig:hyperbolic_equilibria}b), whereas self-propelled tracers
can.  Nevertheless, we show that barriers to active particles still exist.  We also
present theory extending our analysis to the mixing of swimmers in a vortex
flow.

In our model, an ellipsoidal swimmer in two dimensions (2D) is
described by $\bq = (\br,\hat{\bn})$,
comprising its position $\br = (x,y)$ and swimming direction
$\hat{\bn} = (\cos \theta, \sin \theta)$.
Absent noise and active torques, a swimmer with a fixed swimming speed $v_0$ in a fluid velocity field $\bu(\br)$ obeys \cite{Torney2007,Khurana2011,Zottl2012,Arguedas-Leiva2020}
\begin{equation}\label{eq:model_general}
\dot{\br} = \bu + v_0 \hat{\bn}, \quad \quad \quad
\dot{\theta} = \frac{\omega_z}{2}  +\alpha\, \hat{\bn}_\perp \cdot {\bf E} \hat{\bn},
\end{equation}
where $\omega_z = \hat{\bf z} \cdot (\nabla \times \bu)$ is the
vorticity, $\hat{\bn}_\perp = (-\sin \theta,\cos \theta)$,
and ${\bf E} = (\nabla \bu + \nabla \bu^{\rm T})/2$ is the symmetric
rate-of-strain tensor.  The shape parameter $\alpha$ equals 
$(\gamma^2-1)/(\gamma^2+1)$, where 
$\gamma$ is the aspect ratio of the ellipse; $\alpha$  varies from $-1$ to $1$,
where $\alpha = 0$ is a circle, 
and $|\alpha|=1$ is a rod. Positive
(negative) values of $\alpha$ correspond to swimming parallel
(perpendicular) to the major axis. 
The case $\alpha = -1$ coincides with the dynamics of a propagating front element \cite{Mahoney2012} and the optimal (least-time) swimmer trajectories
\cite{Rhoads2013a,ShutingGu2020}.

Equation \eqref{eq:model_general} with $v_0 = 0$ models passive transport.
The linear hyperbolic flow, $\bu = (Ax,-Ay)$ has a passive saddle fixed point at $\br = {\bf 0}$.
The $y$- and $x$-axes are the stable and unstable manifolds, respectively, defined as invariant sets whose points approach
the passive fixed point forwards and backwards in time.
Passive particles cannot cross these passive manifolds (Fig.~\ref{fig:hyperbolic_equilibria}b).

For swimmers in the hyperbolic flow, Eq.~\eqref{eq:model_general}
becomes
\begin{equation}\label{eq:model_hyperbolic}
\dot{\widetilde{x}} = \widetilde{x} + \cos \theta, \quad
\dot{\widetilde{y}} = -\widetilde{y}+ \sin \theta, \quad
\dot{\theta} = -\alpha \sin (2\theta),
\end{equation}
with dimensionless variables $\widetilde{\br}  = (A/v_0)\br$ and 
$\widetilde{t} = A t$.
The natural analogs of the passive fixed point are the fixed points
of Eq.~\eqref{eq:model_hyperbolic}, called \emph{swimming
fixed points} (SFPs) \cite{Berman2020}.  There are four SFPs.
Two SFPs lie on the $y$-axis
with the swimmer facing outward:  $\bq^{\rm out}_{\pm} = (\pm \hat{\bf y}, \pm \hat{\bf y})$.  
The remaining SFPs lie
on the $x$-axis with the swimmer facing inward:  $\bq^{\rm in}_\pm = (\pm \hat{\bf x}, \mp \hat {\bf x})$.  
The SFPs are plotted in
Fig.~\ref{fig:hyperbolic_equilibria}b-d.  These equilibria are saddles,
for all $v_0$ and $\alpha$. 

We set $\alpha = 1$, approximating the shape of 
\emph{B.~subtilis} as a rod.  
Since the SFPs are saddles, they
possess stable and unstable manifolds in the
$\widetilde{x}\widetilde{y}\theta$ phase space, which we call swimming
invariant manifolds (SwIMs) to distinguish them from those for
passive advection.  For $\alpha > 0$, the inward SFPs have two stable
and one unstable direction.  Hence, they each possess a 2D stable SwIM
(Fig.~\ref{fig:hyperbolic_equilibria}c) which together form a warped
sheet in phase space, referred to simply as \emph{the} SwIM.  The SwIM
separates phase space into two regions: to the left [right] of the
SwIM, all swimmer trajectories are ultimately leftward-escaping (LE)
[rightward-escaping (RE)] (Fig.~\ref{fig:hyperbolic_equilibria}d).

The SwIM is only a strict phase-space barrier for \emph{perfectly}
smooth-swimming tracers, which is not the case for real swimmers.
For example, tumbling bacteria apply brief active torques
to suddenly change their swimming direction; we expect these bacteria to be
able to cross the SwIM during their tumbles.  Even for ``smooth-swimming''
bacteria, the swimming direction fluctuates; bacteria wiggle as they swim
due to rotational diffusion \cite{Locsei2009,Junot2019} and
the kinematics of swimming with helical flagella \cite{Hyon2012}.
Hence, bacteria near the SwIM may occasionally cross it due to these small
fluctuations in $\theta$.

The SwIM seen in Fig.~\ref{fig:hyperbolic_equilibria}c produces \emph{one-way} barriers to swimmers when projected into the $\widetilde{x}\widetilde{y}$ plane, barriers that are valid even for noisy swimmers.
For a general 2D flow $\bu(\br)$, a static, parametrized curve $\bR(s)$ with local normal vector $\hat{\bN}(s)$ is a one-way barrier to swimmers when the swimmer velocity across the curve, $[\bu(\bR(s)) + v_0 \hat{\bn}] \cdot \hat{\bN}(s)$, is non-positive for all $\hat{\bn}$.
Hence, if the condition
\begin{equation}\label{eq:oneway}
-\frac{\bu(\bR(s)) \cdot \hat{\bN}(s)}{v_0} \geq 1,\,\,\text{for all } s,
\end{equation}
is met, then the curve $\bR(s)$ is a one-way barrier with local blocking direction $\hat{\bN}(s)$.
For the hyperbolic flow, all non-stationary trajectories along the line $\widetilde{x} = -1$ move leftward, regardless of $\theta$ (Fig.~\ref{fig:hyperbolic_equilibria}d).
Evaluating the left-hand side of Eq.~\eqref{eq:oneway} along this line [in dimensional variables, $\bR(s) = (-v_0/A,s)$ and $\hat{\bN} = \hat{\bf x}$], we obtain identically $1$.
Therefore, this line is a one-way barrier, preventing rightward motion but not leftward.
Furthermore, because Eq.~\eqref{eq:oneway} is independent of $\alpha$ and the time-dependence of $\hat{\bn}$, we expect any curve satisfying it to be a one-way barrier for all swimmers, \emph{regardless of their shape  or motility pattern}.
In particular, we expect the line $\widetilde{x} = -1$ to be a barrier to both the smooth-swimming and tumbling strains of bacteria.

Geometrically, Figs.~\ref{fig:hyperbolic_equilibria}c and \ref{fig:hyperbolic_equilibria}d show that the line $\widetilde{x} = -1$ is the leftmost extent of the 2D SwIM projected into the $\widetilde{x}\widetilde{y}$ plane, i.e.\ it is the left edge of the SwIM.
By symmetry, the right SwIM edge $\widetilde{x} = 1$ is also a one-way barrier, which allows swimmers to pass through it from left to right, but not 
vice-versa.  Hence, the stable SwIM edges form barriers to inward-swimming particles.
Similarly, the horizontal edges of the 2D unstable SwIMs of the
outward SFPs form one-way barriers, blocking outward-swimming particles (Fig.~\ref{fig:hyperbolic_equilibria}b).

We test our theoretical predictions with microfluidic experiments on
swimming bacteria. We fabricate polydimethylsiloxane (PDMS) cells with
channels of width and depth $1\,\,\,{\rm mm}$ in a cross-shaped geometry
(Fig.~\ref{fig:hyperbolic_equilibria}a).  Dilute bacteria
suspensions are pumped into both ends
of the vertical channel and out both ends of the horizontal channel
using syringe pumps.  Microscopy movies are recorded in the center of the channel at 40X. Passive tracer analysis reveals that the flow in the
center (red square in Fig.~\ref{fig:hyperbolic_equilibria}a)
is well-approximated by a 2D linear hyperbolic flow.  The bacteria
used are \emph{B.~subtilis}, either a smooth-swimming strain OI4139 
or a green-fluorescent-protein-expressing 
(GFP) run-and-tumble strain 1A1266.  The bacteria's swimming speeds $v_0$
in the flow have a mean of $25~{\rm \mu m/s}$  and $16~{\rm\mu m/s}$
and standard deviation $11~{\rm \mu m/s}$ and $6~{\rm \mu m/s}$ for the
smooth-swimming and tumbling GFP strains, respectively.  

\begin{figure}
\includegraphics[width=0.5\textwidth]{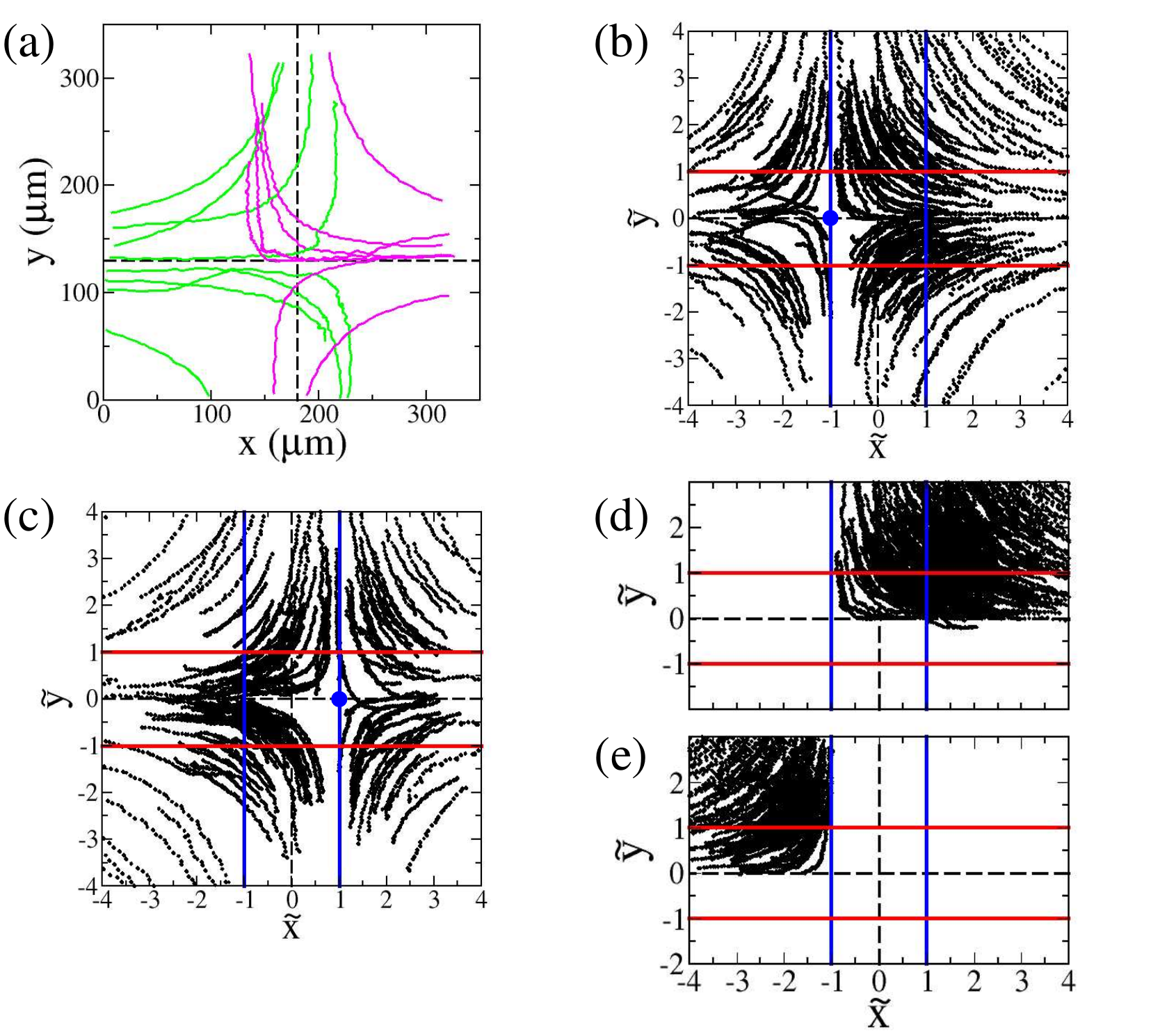}
\caption{(a) Experimental trajectories for smooth-swimming \emph{B.~subtilis};
$A = 0.44\,\,{\rm s}^{-1}$. Passive manifolds are shown with dashed lines.
(b) Right-swimming trajectories.  Positions are scaled by $v_0/A$. 
The theoretically predicted SFP $\bq_-^{\rm in}$ (blue dot) and the SwIM edges (red and blue lines) are shown.
(c) Left-swimming trajectories and $\bq_+^{\rm in}$. (d) Rectified plot showing all trajectories as if leaving
through the upper-right quadrant. (e) All trajectories entering with
$|\widetilde{x}| > 1$ rectified to enter the upper-left quadrant.}
\label{fig:expt_xy_traj}
\end{figure}

Figure \ref{fig:expt_xy_traj} shows  trajectories of smooth-swimming bacteria,
some of which overlap (Fig.~\ref{fig:expt_xy_traj}a).
Trajectories of {\em passive}, non-swimming bacteria in the same experiment
(Supplemental Material Fig.~S1 \cite{SM}) 
are blocked by the vertical passive separatrix (dashed line
in Fig.~\ref{fig:expt_xy_traj}a).
Hence, the region in Fig.~\ref{fig:expt_xy_traj}a where the LE and RE swimmer trajectories
overlap is a signature of the self-propulsion of the swimmers.  Our theory
predicts that the width of this region is the distance between
the vertical SwIM edges shown in Fig.~\ref{fig:hyperbolic_equilibria}b, i.e.\
$2v_0/A$. In the experiments, $v_0$ is approximately constant in time for
individual bacteria; however, different bacteria have different values for
$v_0$ \cite{Kearns2005}. Consequently, the width of the overlap region is undetermined in 
Fig.~\ref{fig:expt_xy_traj}a.

Variations in $v_0$ are accounted for by rescaling 
the spatial coordinates by $v_0/A$,
as in Eq.~\eqref{eq:model_hyperbolic}.
The scaled, non-dimensional trajectories
are shown in
Figs.~\ref{fig:expt_xy_traj}b--e.  The location of the inward SFPs and
their SwIM edges is revealed by plotting trajectories for right-swimming
and left-swimming bacteria separately (Figs.~\ref{fig:expt_xy_traj}b and
\ref{fig:expt_xy_traj}c).  The behavior of inward-swimming bacteria near
an inward SFP is similar to a passive tracer moving near the hyperbolic
fixed point.  
The key difference is that active tracers moving near SFPs can cross the SwIM edge from $\left|\widetilde{x}\right| < 1$ to $\left|\widetilde{x}\right|> 1$, but not in the other direction.

The experimental data are consistent with the theoretically predicted 
one-way barrier property of the SwIM edges.  This is clearest when 
we use the symmetry of Eqs.~\eqref{eq:model_hyperbolic} 
[$(\widetilde{y},\theta) \mapsto (-\widetilde{y},-\theta)$ and 
$(\widetilde{x},\theta) \mapsto (-\widetilde{x},\pi-\theta)$] to
rectify the trajectories, such that all trajectories are displayed
as though entering from the upper inlet and escaping to the right.  
Under this transformation, Fig.~\ref{fig:expt_xy_traj}d shows that 
all trajectories are bounded from the left by the SwIM edge at 
$\widetilde{x} = -1$, in agreement with the theory.  Indeed, any 
bacterium crossing this SwIM edge from left to right would violate 
the one-way barrier property.
Furthermore, all bacteria that enter with $\left| \widetilde{x} \right| >
1$ (Fig.~\ref{fig:expt_xy_traj}e, rectified such that initial 
$\widetilde{x} < -1$) are swept away from the center of the
cell, consistent with the SwIM edges at $\left| \widetilde{x} \right| =1$ as barriers to inward-swimming bacteria.

\begin{figure}
\centering
\includegraphics[width=0.5\textwidth]{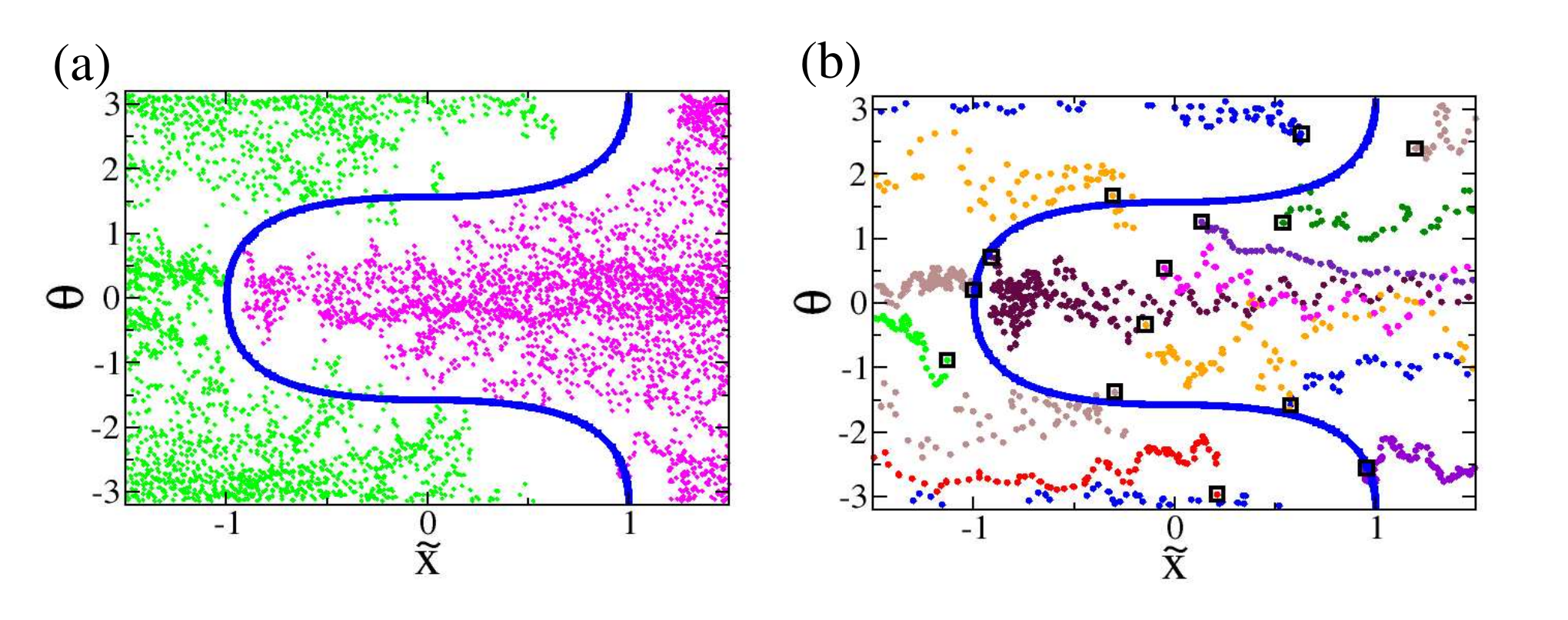}
\caption{Experimental $\widetilde{x} \theta$ 
 trajectories for smooth-swimming \emph{B.~subtilis}; 
$A = 0.44\,\,\,{\rm s}^{-1}$. The theoretical SwIM ($\alpha = 1$) is plotted in
blue.  (a) All trajectories. Leftward-escaping trajectories are 
green, and rightward-escaping trajectories are magenta. (b) Selected
trajectories; the beginning of each is marked with an open
square.} 
\label{fig:x_theta}
\end{figure}
The delineation between LE and RE swimmers by the SwIM in the
$\widetilde{x}\theta$ plane is shown experimentally in Fig.~\ref{fig:x_theta} (see \cite{SM} for the measurement of $\theta$).
Most of the trajectories in Fig.~\ref{fig:x_theta}a respect this
barrier, although there is a slight breach of the SwIM for some of the
bacteria, due to the variations in $\theta$ discussed previously.
These vertical fluctuations in individual trajectories
(Fig.~\ref{fig:x_theta}b) cause momentary crossings of the ``horizontal" part of the SwIM.

%
\begin{figure}
\includegraphics[width=0.5\textwidth]{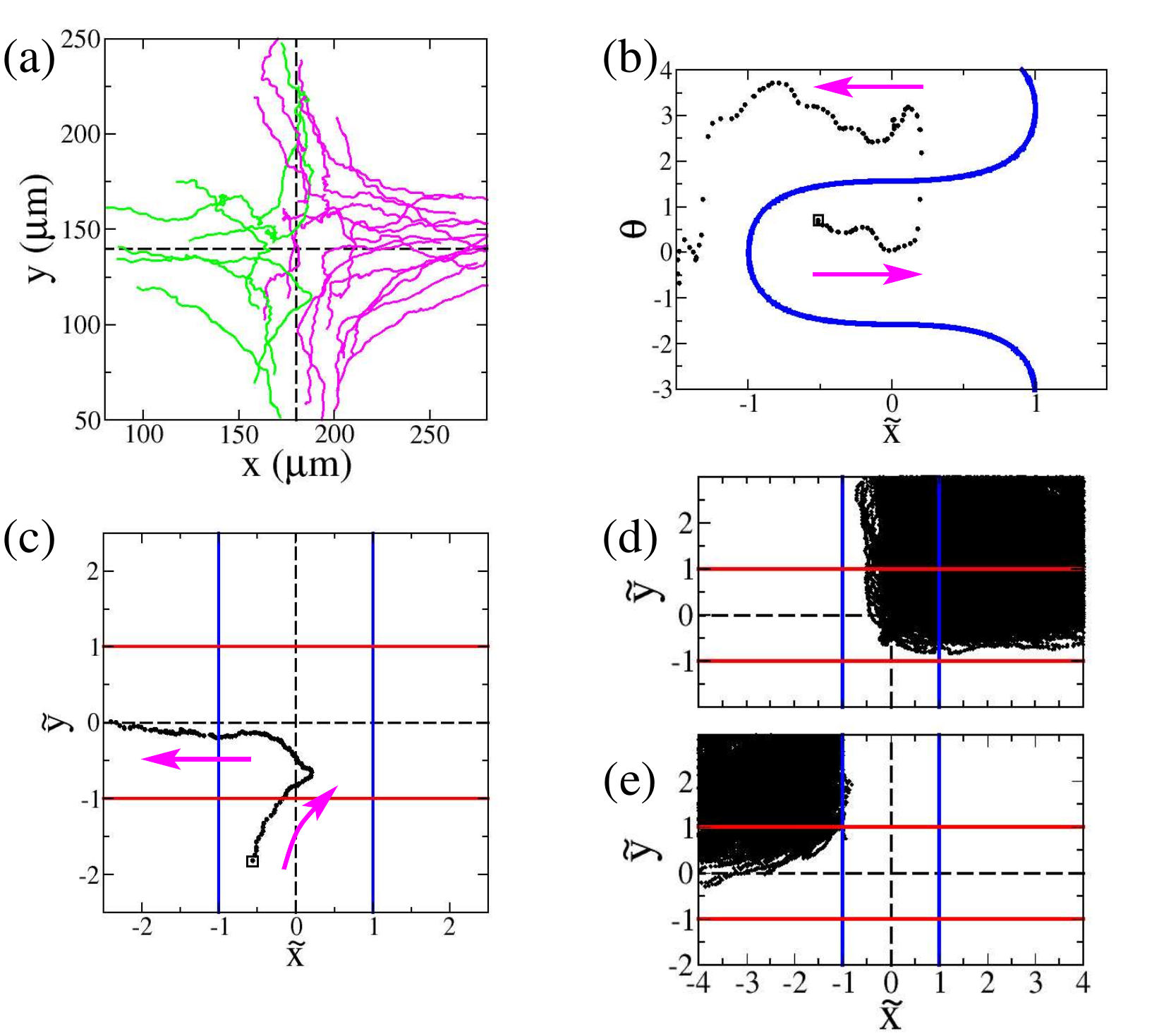}
\caption{(a) Selected trajectories of run-and-tumble \emph{B.~subtilis}; $A = 0.44\,\,{\rm s}^{-1}$. Passive
  manifolds are shown with dashed lines.  (b) $\widetilde{x} \theta$
  plot and (c) $\widetilde{x}\widetilde{y}$ trajectory for a single
  bacterium with well-defined tumbling events. (d) and (e) Scaled and
  rectified trajectories for tumbling bacteria, as in
  Figs.~\ref{fig:expt_xy_traj}d and
  \ref{fig:expt_xy_traj}e.}\label{fig:gfp}
\end{figure}
Angular fluctuations are, of course, particularly pronounced for the 
tumbling strain of bacteria (Fig.~\ref{fig:gfp}a), leading to highly 
irregular $\widetilde{x}\theta$ trajectories.
However, for bacteria with well-defined tumble events, the
$\widetilde{x}\theta$ trajectories (Fig.~\ref{fig:gfp}b) give
insight into the short-term direction (right or left) of their 
$\widetilde{x}\widetilde{y}$ motion (Fig.~\ref{fig:gfp}c). The bacterium in
these two plots begins to the right of the SwIM; the corresponding $\widetilde{x}\widetilde{y}$ trajectory moves
to the right during this period. The bacterium undergoes a significant tumble at $\widetilde{x} = 0.2$, jumping above and to the
left of the SwIM (Fig.~\ref{fig:gfp}b), with a corresponding change in
direction in the $\widetilde{x}\widetilde{y}$ plane (Fig.~\ref{fig:gfp}c).

Despite the dramatic fluctuations in their orientations, the tumbling
bacteria's $\widetilde{x}\widetilde{y}$ trajectories
respect the vertical lines $\widetilde{x} = \pm 1$ as one-way
barriers, as predicted.  Any RE swimmer must have entered with
$\widetilde{x} > -1$ (Fig.~\ref{fig:gfp}d), and any swimmer that
enters with $\widetilde{x} < -1$ must move leftward, away from the
SwIM edge (Fig.~\ref{fig:gfp}e).
Furthermore, though the trajectories in Fig.~\ref{fig:gfp}d
cross the horizontal passive manifold, they do not cross the lower red
line at $\widetilde{y} = -1$, respecting its outward-blocking nature.

In arbitrary flows, SwIM edges may not act as barriers for tumbling bacteria because they do not satisfy Eq.~\eqref{eq:oneway} in general.
However, BIMs---which were introduced as one-way barriers to \emph{front} propagation---always satisfy Eq.~\eqref{eq:oneway}.
In 2D time-independent flows, BIMs are the one-dimensional SwIMs for the $\alpha = -1$ case of Eq.~\eqref{eq:model_general} (i.e.\ $\alpha = -1$ trajectories $\bq(t)$ that are asymptotic to SFPs), which satisfy the condition $-\bu(\br(t)) \cdot \hat{\bn}(t)/v_0 = 1$  \cite{Mitchell2012,Mahoney2015,SM}.
Therefore, we now recognize BIMs as one-way barriers for all swimmers of a fixed swimming speed $v_0$, including those exhibiting rotational diffusion, tumbling, or other reorientation mechanisms.
The robust bounding behavior  occurs in our experiments because the SwIM edges coincide with the BIMs for linear hyperbolic flows. 
In general nonlinear flows, SwIM edges and BIMs depart from each other. 
Thus, the SwIM edges are the more relevant barriers for
perfect smooth swimmers, whereas the BIMs are more relevant for noisy
swimmers, as we illustrate with the following example.


%
\begin{figure}
\includegraphics[width=0.47\textwidth]{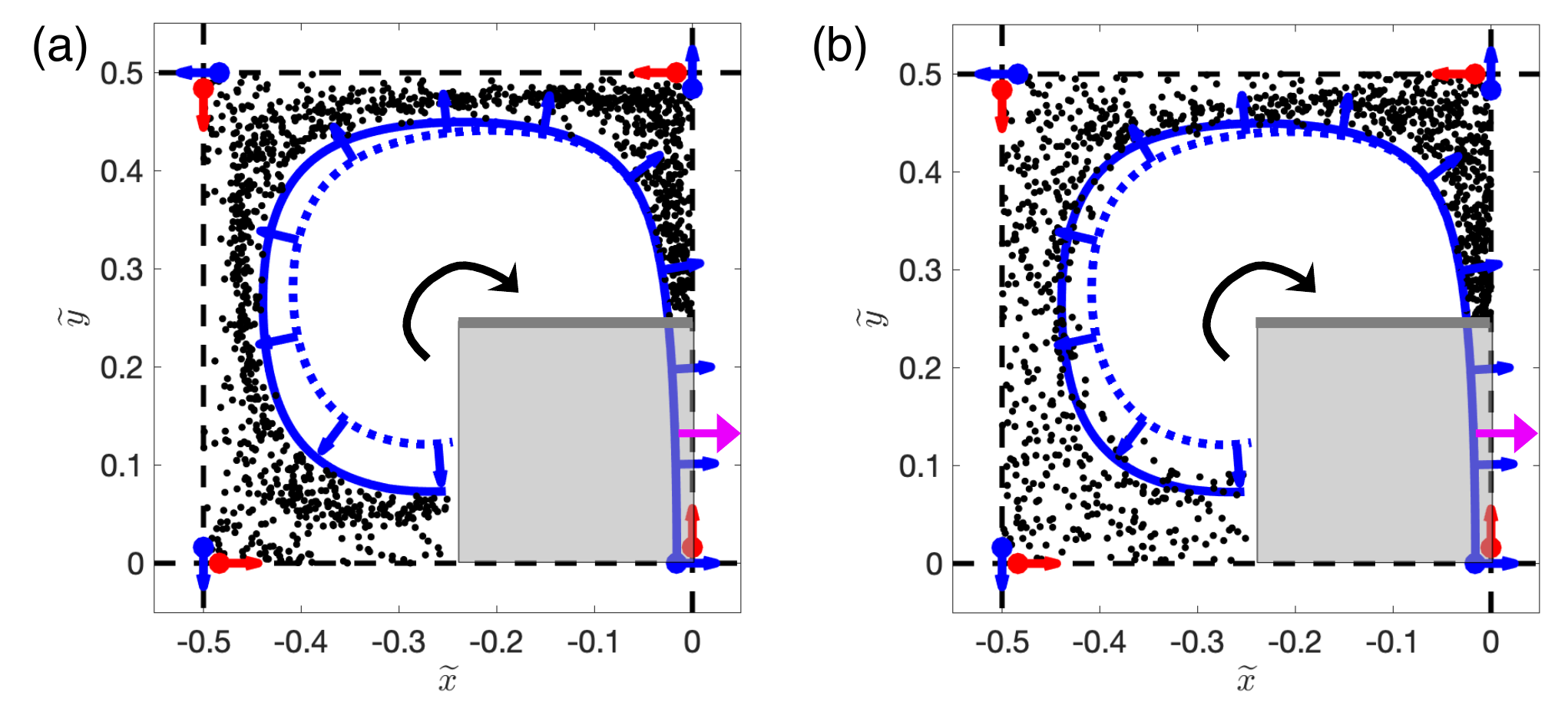}
\caption{Bounding properties of SwIM edges and BIMs in a vortex flow; $v_0/U = 0.1$, $\alpha =1$.
(a) Initial positions (black dots)
of smooth-swimmers that enter the grey square from the upper side and exit
it on the right side (magenta arrow). 
The stable SwIM edge and stable BIM of the lower-right SFP 
are shown as solid blue and dotted blue curves, respectively.
(b) Same as panel (a) for swimmers with rotational diffusivity $D_r$; 
$D_r L/U = 0.86$.
}
\label{fig:vortex}
\end{figure}
We consider the swimmer dynamics Eq.~\eqref{eq:model_general} in the vortex-lattice flow \cite{Torney2007,Khurana2011,Ariel2017,Berman2020} 
 $\bu = (\sin (2\pi \widetilde{x}) \cos(2\pi \widetilde{y}),-\cos(2\pi \widetilde{x})\sin(2\pi \widetilde{y}))$, where we use non-dimensional coordinates $\widetilde{\br} = \br/L$ and $\widetilde{t} = tU/L$
for a flow with maximum speed $U$ and length scale $L$. 
Near $\br=0$, the flow is approximately the linear hyperbolic 
flow, with  $A = 2\pi$.  Thus, the origin is surrounded by SFPs 
(Fig.~\ref{fig:vortex}a)
analogous to those
of Eq.~\eqref{eq:model_hyperbolic} \cite{Berman2020}.  

In analogy with the preceding microfluidic experiments that identified
the positions of RE trajectories, we perform the following numerical
experiment.  We integrate the initial conditions of swimmers selected
at random inside a single vortex cell but outside the grey square
shown in Fig.~\ref{fig:vortex}a.  We then plot only those initial
positions for which the swimmer trajectory enters the grey square at
the upper edge $\widetilde{y} = 0.25$ and subsequently exits through
the right edge at $\widetilde{x} = 0$ (see \cite{SM} for animations).  These trajectories are
analogous to the RE trajectories in the experimental hyperbolic flow.
Figure~\ref{fig:vortex}a shows the  result of the calculation for
perfect smooth swimmers, along with the SwIM edge for the 2D stable
SwIM of the vortex flow (solid curve) and the corresponding BIM
(dotted curve).  Clearly, these initial conditions are bounded by the
SwIM edge, showing that the SwIM edge again bounds those trajectories
that exit right, even in a nonlinear flow.
We repeat the calculation with a moderate-intensity white noise term added to $\dot{\theta}$ in Eq.~\eqref{eq:model_general} to simulate rotational diffusion for realistic smooth-swimming bacteria  \cite{Locsei2009}.
The resulting set of initial conditions (Fig.~\ref{fig:vortex}b) breaches the SwIM edge, but it remains bounded by the BIM, consistent with the absolute one-way barrier property of BIMs for all swimmers, regardless of their reorientation mechanism.

In summary, we have shown theoretically and experimentally that the
trajectories of self-propelled particles in externally-driven fluid
flows are constrained by the presence of one-way barriers, i.e.\ SwIM
edges and BIMs.  Despite the simplicity of our model, we are able
to fully explain certain properties of the trajectories of swimming
bacteria in an externally-driven microfluidic flow. 
Our SwIM framework provides a foundation for understanding the critical barrier structures
that dominate the mixing of a wide range of self-propelled tracers
in laminar flows. 
For example, BIMs must also block gyrotactic \cite{Guasto2012,Cencini2019}  and chemotactic swimmers, since these barriers are independent of biases on the swimming direction.
We further expect that the SwIM approach can be generalized to more complicated, time-periodic,
time-aperiodic and weakly turbulent flows. 
It remains an open question how our approach may apply to the trajectories
of self-propelled agents in active matter systems featuring self-driven
flows, such as individual bacteria within a swarm \cite{Ariel2015}
or motile defects in active nematics \cite{Sanchez2012,Shankar2018,Tan2019}.

\begin{acknowledgements}
These studies were supported by the National Science Foundation under grants
DMR-1806355 and CMMI-1825379. We thank Nico Waisbord and Jeff Guasto for
providing the smooth-swimmer strain used in these experiments,
Jack Raup and Joe Tolman for assistance
with milling, Matt Heinzelmann for assistance with the incubation techniques,
and Brandon Vogel for guidance on PDMS techniques.
\end{acknowledgements}

\end{document}


\title{Transport barriers to self-propelled particles in fluid flows: Supplemental Material}
\author{Simon A. Berman, John Buggeln, David Brantley,  Kevin A. Mitchell,\\ Thomas H. Solomon}
\date{}
\maketitle

\section{Experimental methods}

\subsection{Incubation protocols} 

For the tumbling, GFP bacteria (1A1266),
a scraping from a sample frozen at $-80^{\circ}$C is used to inoculate
50 mL of LB broth. This broth is incubated
overnight at $37^{\circ}$C (with shaking), reinoculated in the morning
in fresh LB broth, incubated for 3 hours, and then diluted to an optical
density (OD)$_{600} \sim 0.02$. (A shorter incubation time results in a
smaller optical density without need for dilution, but the bacteria rely
on quorum sensing to turn on their swimming, so that approach results
in most of the bacteria remaining sessile.)

The smooth-swimming bacteria (OI4139) are incubated in 
 CAP medium \cite{Rusconi2014}. The following stock solutions are used:
(a) a salt solution, composed of 6.80 g KH$_2$PO$_4$, 8.70 g K$_2$HPO$_4$, 
0.240 g MgCl$_2$, 0.132 g NH$_4$SO$_4$, and 0.00125 g MnCl$_2$, mixed in
1.0 L of distilled water and autoclaved (a precipitate forms during
autoclaving, which is filtered out later in the process); (b) 1.00 g of CaCl$_2$ in 50 ml
of water; (c) 0.050 g each of histidine, methionine, and tryptophan in 10
ml of water; (d) 1.80 g of sorbitol in 10 ml of water; (e) 10.0 g of
tryptone and 5.0 g NaCl in 1 L of water (or 15.0 g of ``tryptone
water'' powder in 1 L of water); (f) 1.20 g of IPTG in 20 ml of water; and
(g) 2.5  ml of glycerin in 47.5 ml of water.  To prepare incubation media,
50 ml of
the salt solution is combined with  0.23 ml of the CaCl$_2$ solution, 0.15
ml of the HMT solution, 0.50 ml each of the sorbitol and Tryptone broth
solutions, and 0.20 ml of IPTG solution.  (For the
overnight incubations, we often double the tryptone broth concentration.)
This mixture is then filter-sterilized and divided into ten 
5-ml samples, each in a 15-ml capacity centrifuge tube.  

Scrapings of smooth-swimming bacteria (OI4139) from frozen stock are used to
inoculate one of the 5-ml CAP solutions, which is then incubated overnight
at $37^{\circ}$C (with shaking), reinoculated in the morning into a fresh
CAP solution, incubated for 4 hours, injected with 0.050 ml (2 drops) of 
the glycerol solution,
incubated for an additional 15 minutes, and then diluted to an 
optical density (OD)$_{600} \sim 0.03$. 

\subsection{Tracking and three-dimensional (3D) effects} 

Trajectories for both passive microspheres and bacteria 
are extracted using
standard particle tracking algorithms \cite{Crocker1996}.
The swimming speed $v_0$ and the orientation $\theta$ of each bacterium
are determined by subtracting the hyperbolic flow velocity from the velocity
of the trajectory: $v_0 \hat{\bn} = \dot{\br} - \bu(\br)$ \cite{Barry2015,Junot2019}.

A small fraction of the smooth-swimming bacteria still tumble
after these procedures. We identify those bacteria from the
trajectories by large, abrupt changes in orientation angle $\theta$; 
these tumblers are discarded from our analysis of the smooth-swimming
bacteria.

The analysis treats the bacteria as though swimming in a two-dimensional (2D)
plane. However, even though the flow itself is predominantly 2D, the 
bacteria can swim with velocity components perpendicular to the focal
plane of the microscope.
With the 40X objective, the focal plane of the microscope is able to
image bacteria
over a thickness of approximately 20 - 30 $\mu$m for the smooth swimming
bacteria using diascopic microscopy, and approximately 50 - 60 $\mu$m for
the fluorescent (tumbling) bacteria imaged under epifluorescence illumination.
Given typical tracking times and swimming speeds, the smooth swimming bacteria
tracked in Fig.~2 likely have swimming orientations that are
within $20^{\circ}$ of the 2D focal plane. Of course, the 
tumbling bacteria in Fig.~4 may have orientations at 
larger angles relative to the focal plane, especially right after a tumble
event. In those cases, the bacteria would swim out of the focal region
quickly, so those larger angles are presumably near the ends of the
plotted trajectories. 

Ultimately, the 2D trajectories plotted and analyzed are 2D 
projections of the full 3D motion. The true swimming speed of the bacteria
may be larger than that reported (which is based on the 2D projection),
but the theory accurately captures the behavior of the 2D projections of
the trajectories, based on the swimming speeds of the projections.

\section{Additional experimental data}

\subsection{Passive (Sessile) bacteria}

\begin{figure}
\centering
\includegraphics[width=0.8\textwidth]{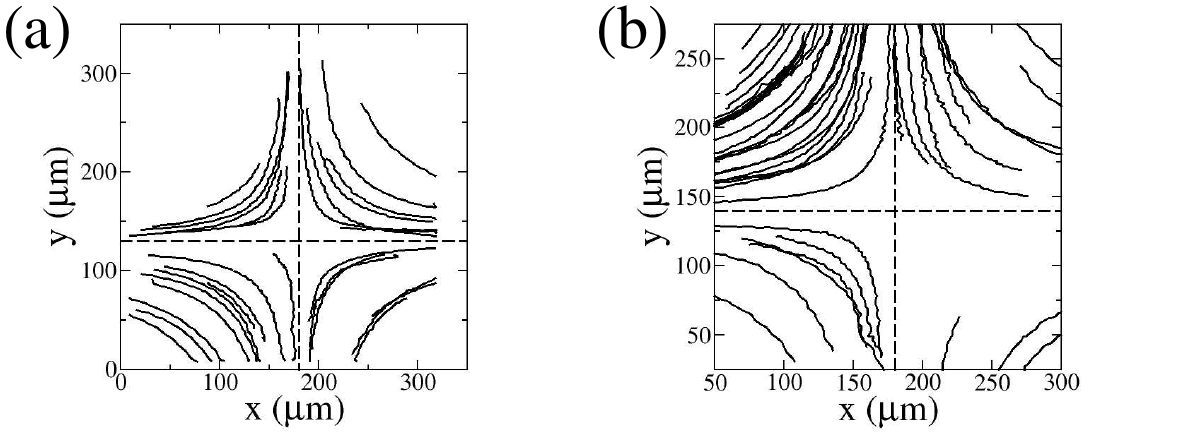}
\caption{Trajectories of sessile bacteria in a flow with $A = 0.44$ s$^{-1}$.  (a) Sessile bacteria in
the same experiment as shown in Fig. 2a.
(b) Sessile bacteria in the same experiment as shown in Fig. 4a. The dashed
lines show the same passive manifolds as in Figs. 2 and 4.}
\label{fig:supp_sessile}
\end{figure}

For each of the experimental runs reported, there are bacteria that are
sessile (non-motile) or that swim with sufficiently small speeds as to behave
as though sessile.  We plot the trajectories of these sessile bacteria in
Fig.~\ref{fig:supp_sessile}. For these non-swimming bacteria, the
passive manifolds are the barriers. This verifies that the crossing
of the passive manifold seen in Figs.~2 and 4 is due 
to the swimming of the organisms.

\subsection{Motile bacteria in a slower flow}

\begin{figure}
\centering
\includegraphics[width=0.8\textwidth]{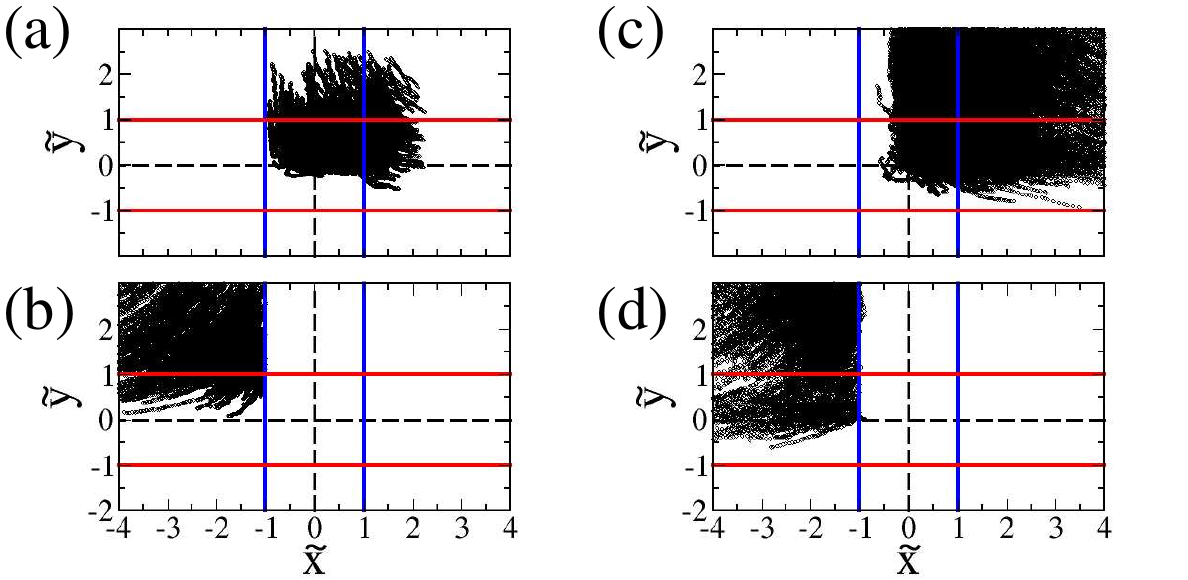}
\caption{Scaled bacteria trajectories ($\widetilde{x} = (A/v_0)x$)
for a hyperbolic flow  with $A = 0.19\,\,\, $s$^{-1}$. The left two panels
show results for smooth-swimming bacteria: (a) rectified plot with every
trajectory that leaves to the right; and (b) rectified plot with every
trajectory that enters with non-dimensional $\widetilde{x} < -1.0$. The right
two panels show similar data for tumbling bacteria.  Passive manifolds
are shown with dashed lines and the theoretically predicted SwIM edges
are shown with red and blue lines.
}
\label{fig:supp_003}
\end{figure}

The data in Figs.~2 and 4 cover a range of $v_0/A$ of over
a factor of two in each case, since the bacteria swim with a range of
swimming speeds. But the theory works equally well if $v_0/A$ is varied
by changing the flow speed parameter $A$. Figure~\ref{fig:supp_003} shows data similar
to that in Figs. 2 and 4, but for a flow rate less than half as strong.
Once again, every trajectory that ends up leaving toward the right
enters to the right of the left SwIM edge (Figs.~\ref{fig:supp_003}a and \ref{fig:supp_003}c). 
And every trajectory that enters to the left of the left SwIM edge exits 
to the left (Figs.~\ref{fig:supp_003}b and \ref{fig:supp_003}d).

There is a depletion zone next to the SwIM edge for the tumbling bacteria
(Fig.~\ref{fig:supp_003}c), which is also seen for the higher flow rate
(Fig. 4d). This depletion region exists because tumbling
bacteria that begin on the right of the SwIM edge can tumble and swim left across this
one-way barrier, but they can't cross back once outside the SwIM edge, since
it blocks inward-swimming bacteria.

\section{Burning invariant manifolds (BIMs)}
Here we provide background information on burning invariant manifold (BIM) theory \cite{Mahoney2012,Mitchell2012,Mahoney2015}.
Initially, this theory was developed to describe barriers to the propagation of chemical reaction fronts in a 2D fluid flow $\bu(\br)$.
Fronts are represented by parametrized curves $(\br(\lambda),\theta(\lambda))$, where $\br$ is the 2D position and $\theta$ is the angle of the normal to the front $\hat{\bn}(\lambda) = \cos\theta(\lambda) \hat{\bf x} + \sin\theta(\lambda) \hat{\bf y}$, which satisfy
\begin{equation}\label{eq:frontcompat}
\frac{d \br}{d \lambda} \cdot \hat{\bn}(\lambda) = 0.
\end{equation}
Equation \eqref{eq:frontcompat} is known as the front-compatibility criterion \cite{Mitchell2012}.
A point on the front, i.e.\ a front element $(\br,\theta)$, evolves in time according to the equations
\begin{subequations}\label{eq:frontelem}
\begin{align}
\dot{\br} & = \bu(\br) + v_0 \hat{\bn}, \\
\dot{\theta} & = \frac{\omega_z}{2}  - \, \hat{\bn}_\perp \cdot {\bf E} \hat{\bn},
\end{align}
\end{subequations}
where $v_0$ is the local front-propagation speed and $\omega_z$ and ${\bf E}$ are the vorticity and rate-of-strain tensor, respectively.
Equation \eqref{eq:frontelem} is simply Eq.~(1) with $\alpha = -1$; hence, front-element dynamics is equivalent to the dynamics of a rod-shaped swimmer that swims perpendicular to its axis \cite{Mitchell2012}.

BIMs are defined as the one-dimensional invariant manifolds of the fixed points of Eq.~\eqref{eq:frontelem}.
Because Eq.~\eqref{eq:frontelem} assumes the flow has no explicit time-dependence, BIMs are solutions of Eq.~\eqref{eq:frontelem} that asymptotically approach a fixed point forwards or backwards in time.
They are special cases of swimming invariant manifolds (SwIMs): they are the one-dimensional SwIMs of the swimming fixed points (SFPs) of $\alpha = -1$ swimmers. 
For example, for the hyperbolic flow [Eq.~(2)], the SFPs $\bq_\pm^{\rm in}$ have one stable and two unstable directions when $\alpha = -1$.
Thus, their one-dimensional stable manifolds are BIMs (blue curves in Fig. 1b).
Conversely, the SFPs $\bq_\pm^{\rm out}$ have one unstable and two stable directions when $\alpha = -1$.
Thus, their one-dimensional unstable manifolds are BIMs (red curves in Fig. 1b).
For the vortex flow, the SFPs surrounding the hyperbolic fixed points also possess BIMs, i.e.\ one-dimensional invariant manifolds when $\alpha = -1$, so long as $v_0$ is not too large \cite{Berman2020}.
 
BIMs belong to a special family of solutions of Eq.~\eqref{eq:frontelem}  for which the front element trajectories are themselves fronts.
These special solutions, called sliding fronts \cite{Mahoney2015}, are defined as those satisfying Eq.~\eqref{eq:frontcompat} with $\lambda \rightarrow t$.
Explicitly, we have
\begin{equation}\label{eq:slidfront}
\dot{\br}(t) \cdot \hat{\bn}(t) = \bu(\br(t)) \cdot \hat{\bn}(t) + v_0= 0
\end{equation}
all along the trajectory, which is known as the sliding-front condition.
It follows from direct calculation that if Eq.~\eqref{eq:slidfront} is satisfied at some initial time, then it is satisfied for all time \cite{Mahoney2015}.
In particular, this is true for BIMs, and hence the BIM trajectory $(\br(t),\theta(t))$ is itself a front.
By virtue of Eq.~\eqref{eq:slidfront}, BIMs satisfy the inequality Eq.~(3).
Therefore, they are one-way barriers to swimmers, regardless of $\alpha$ or any noise in the orientation.

\section{Vortex flow simulations}
%
To simulate the swimmer dynamics in the vortex flow, we investigate the trajectories of Eq.~(1) with  $\bu = (U\sin (2\pi x/L) \cos(2\pi y/L),-U\cos(2\pi x/L)\sin(2\pi y/L))$, where $L$ is the flow length scale and $U$ is the maximum flow speed.
Thus, the equations of motion are
\begin{subequations}\label{eq:model_vortex}
\begin{align} \label{eq:xdot}
\dot{x} & = U\sin \left(\frac{2 \pi x}{L}\right) \cos \left(\frac{2 \pi y}{L}\right) + v_0 \cos \theta, \\ \label{eq:ydot}
\dot{y} & = -U\cos \left(\frac{2 \pi x}{L}\right) \sin \left(\frac{2 \pi y}{L}\right)+v_0 \sin \theta, \\ \label{eq:thdot}
\dot{\theta} & = \frac{2\pi U}{L}\left[ \sin \left(\frac{2 \pi x}{L}\right) \sin \left(\frac{2 \pi y}{L}\right) - \alpha \cos\left(\frac{2 \pi x}{L}\right) \cos\left(\frac{2 \pi y}{L}\right) \sin (2\theta) \right] + \xi(t).
\end{align}
\end{subequations}
Here, we have added a Gaussian white-noise term $\xi(t)$ to Eq.~\eqref{eq:thdot} to model rotational diffusion.
The noise is zero-mean and delta-correlated with rotational diffusivity $D_r$:
\begin{subequations}
\begin{align}
\langle \xi(t) \rangle & = 0, \\
\langle \xi(t) \xi(t') \rangle & = 2 D_r \delta(t- t').
\end{align}
\end{subequations}
Non-dimensionalizing Eq.~\eqref{eq:model_vortex} with coordinates $\widetilde{\br} = \br/L$ and $\widetilde{t} = tU/L$, we find that the relevant system parameters are the dimensionless quantities $v_0/U$, $D_r L/U$, and $\alpha$.
All calculations in the main text are for $\alpha = 1$ swimmers.

\begin{table}
\centering
\begin{tabular}{l | c}
Event & Symbol \\ \hline
Moving down through $\widetilde{y} = 0.25$ & 1 \\
Moving right through $\widetilde{x}=0$ & 2 \\
Moving down through $\widetilde{y} = 0$ & 3 \\
Moving left through $\widetilde{x} = -0.25$ & 4 \\
Moving left through $\widetilde{x} = -0.5$ & 5 \\
Moving up through $\widetilde{y} = 0.5$ & 6
\end{tabular}
\caption{Symbols assigned to swimmer trajectories for certain events.}\label{tab:symbols}
\end{table}
We select $N$ random swimmer initial conditions and integrate them from $\widetilde{t} = 0$ to $\widetilde{t} = 3$.
The initial positions are uniformly randomly distributed inside a single vortex cell with $-0.5 < \widetilde{x} < 0$ and $0 < \widetilde{y} < 0.5$ but outside the grey square illustrated in Fig.~5, defined by $-0.25 \leq  \widetilde{x} \leq 0$ and $0 \leq  \widetilde{y} \leq 0.25$, and the initial orientations are also uniformly randomly distributed in the full range $\theta \in [0,2\pi]$.
For each trajectory, we record certain events related to entering or leaving the grey square or the vortex cell using a numeric symbol, as detailed in Table \ref{tab:symbols}.
Clearly, a swimmer must register one of the symbols ``2," ``3," ``5," or ``6" when it exits the vortex cell.
For a swimmer that registers a symbol ``1" immediately followed by ``2," ``3," or ``4," it is assumed the swimmer entered the square through the top side and subsequently exited via one of the remaining sides.
For instance, the label ``12" means the swimmer enters the square from the upper side, exits on the right side, and never subsequently returns to the square; the label ``1413" means the swimmer completes a full revolution around the vortex center, returns to the square and exits on the lower side.

In Fig.~5, we plot only the initial positions of swimmers whose trajectory labels begin with ``12," from a simulation with $N  \approx 2 \times 10^4$.
In Supplemental Movie S1, we provide an animation of the calculation with $D_r=0$ leading to Fig.~5a, i.e.\ for perfect smooth swimmers (with $N  \approx 10^3$ for clarity).
Swimmers whose labels begin with ``12" are the large, green dots, while the remaining swimmers whose first symbols are ``1" are the small, grey dots.
The arrows indicate instantaneous swimming directions.
All other trajectories are discarded from the animation for clarity.
In Supplemental Movie S2, we provide an animation of the analogous calculation with $D_rL/U = 0.86$ leading to Fig.~5b.

\begin{figure}
\centering
\includegraphics[width=0.8\textwidth]{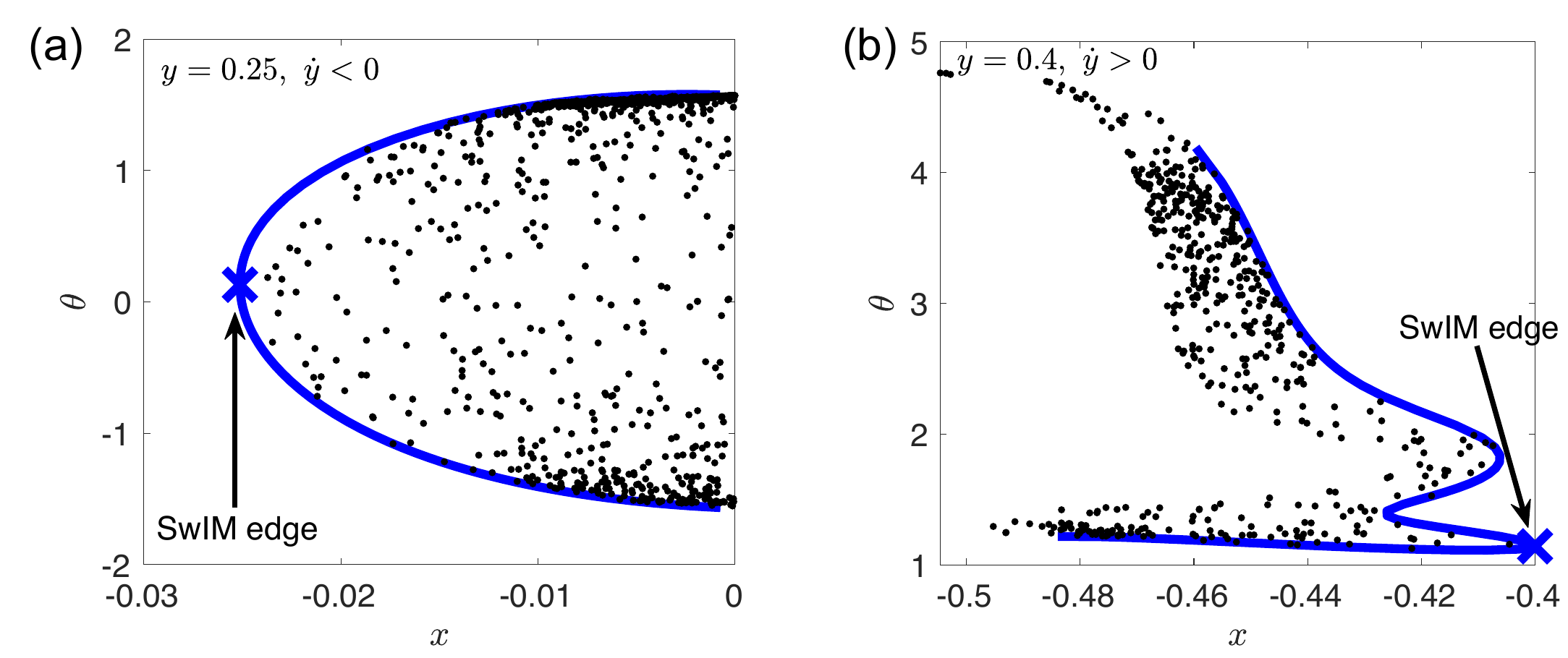}
\caption{Cross sections of stable SwIM (blue curves) at $\widetilde{y} = 0.25$ [$\dot{\widetilde{y}} < 0$, (a)] and $\widetilde{y} = 0.4$  [$\dot{\widetilde{y}} > 0$, (b)]. The black dots show the intersections of the ``12" swimmer trajectories, which eventually exit right, with the surface of section; $D_r = 0$.}\label{fig:theory1}
\end{figure}

For the vortex flow \eqref{eq:model_vortex}, we compute the SwIM edge as follows.
Equations \eqref{eq:model_vortex} have an SFP at $\bq = (-\sin^{-1}(v_0/U)/(2\pi),0,0)$ in rescaled units, which is a saddle with two stable directions and one unstable \cite{Berman2020}.
We compute the 2D stable SwIM by integrating trajectories initialized near the SFP in the 2D stable subspace backwards in time.
Cross-sections of part of the 2D SwIM at different $\widetilde{y}$ positions away from the SFP are shown in Fig.~\ref{fig:theory1}.
We locate the SwIM edge at every angle $\phi$ relative to the center of the vortex as the point on the SwIM which is closest to the vortex center at that $\phi$.
These points are indicated in Fig.~\ref{fig:theory1} by the arrows.
Figure \ref{fig:theory1} shows that all swimmers that exit right remain on the outer side of the SwIM (relative to the vortex center).
Hence, in $\widetilde{x}\widetilde{y}$ space, their trajectories are bounded by the SwIM edge.